\documentclass[manuscript]{acmart}
\AtBeginDocument{%
  \providecommand\BibTeX{{%
    \normalfont B\kern-0.5em{\scshape i\kern-0.25em b}\kern-0.8em\TeX}}}

\copyrightyear{2024}
\acmYear{2024}
\setcopyright{rightsretained}
\acmConference[CHI '24]{Proceedings of the CHI Conference on Human Factors in Computing Systems}{May 11--16, 2024}{Honolulu, HI, USA}
\acmBooktitle{Proceedings of the CHI Conference on Human Factors in Computing Systems (CHI '24), May 11--16, 2024, Honolulu, HI, USA}
\acmDOI{10.1145/3613904.3642233}
\acmISBN{979-8-4007-0330-0/24/05}

\newcommand{\red}[1]{\textcolor{black}{#1}}

\begin{document}



\title{A Contextual Inquiry of People with Vision Impairments in Cooking}


\author{Franklin Mingzhe Li}
\affiliation{%
  \institution{Carnegie Mellon University}
  \city{Pittsburgh}
  \state{PA}
  \country{United States}
}
\email{mingzhe2@cs.cmu.edu}

\author{Michael Xieyang Liu}
\affiliation{%
  \institution{Google Research}
  \city{Mountain View}
  \state{CA}
  \country{United States}
}
\email{lxieyang@google.com}

\author{Shaun K. Kane}
\affiliation{%
  \institution{Google Research}
  \city{Boulder}
  \state{Colorado}
  \country{United States}
}
\email{shaunkane@google.com}

\author{Patrick Carrington}
\affiliation{%
  \institution{Carnegie Mellon University}
  \city{Pittsburgh}
  \state{PA}
  \country{United States}
}
\email{pcarrington@cmu.edu}
\renewcommand{\shortauthors}{Li et al.}



\begin{abstract}
Individuals with vision impairments employ a variety of strategies for object identification, such as pans or soy sauce, in the culinary process. In addition, they often rely on contextual details about objects, such as location, orientation, and current status, to autonomously execute cooking activities. To understand how people with vision impairments collect and use the contextual information of objects while cooking, we conducted a contextual inquiry study with 12 participants in their own kitchens. This research aims to analyze object interaction dynamics in culinary practices to enhance assistive vision technologies for visually impaired cooks. We outline eight different types of contextual information and the strategies that blind cooks currently use to access the information while preparing meals. Further, we discuss preferences for communicating contextual information about kitchen objects as well as considerations for the deployment of AI-powered assistive technologies.
\end{abstract}




\maketitle

\section{Introduction}
Cooking holds a profound sway over the overall quality of life for individuals with vision impairments \cite{li2021non,bilyk2009food}. 
Nevertheless, the culinary process leans heavily on visual cues, relying extensively on the contextual information of objects nestled within the kitchen (e.g., location, status), which remains elusive to people with vision impairments \cite{li2021non,tian2010improving}. This creates substantial barriers for them to cook independently and, ultimately, negatively impacts their quality of life \cite{jones2019analysis}.
For example, seemingly simple tasks like finding an item in a cluttered refrigerator or gauging whether a dish is fully cooked are challenging for visually impaired individuals, particularly for beginners in rehabilitation training programs \cite{li2021non,wang2023practices}. Prior research has highlighted the importance of recognizing contextual information \cite{tian2010improving}, such as the color and placement of objects in everyday tasks. However, there exists a notable gap of knowledge regarding the specific contextual cues that visually impaired people rely on and the rationale behind these preferences when cooking.\looseness=-1

\begin{figure*}
  \includegraphics[width=\textwidth]{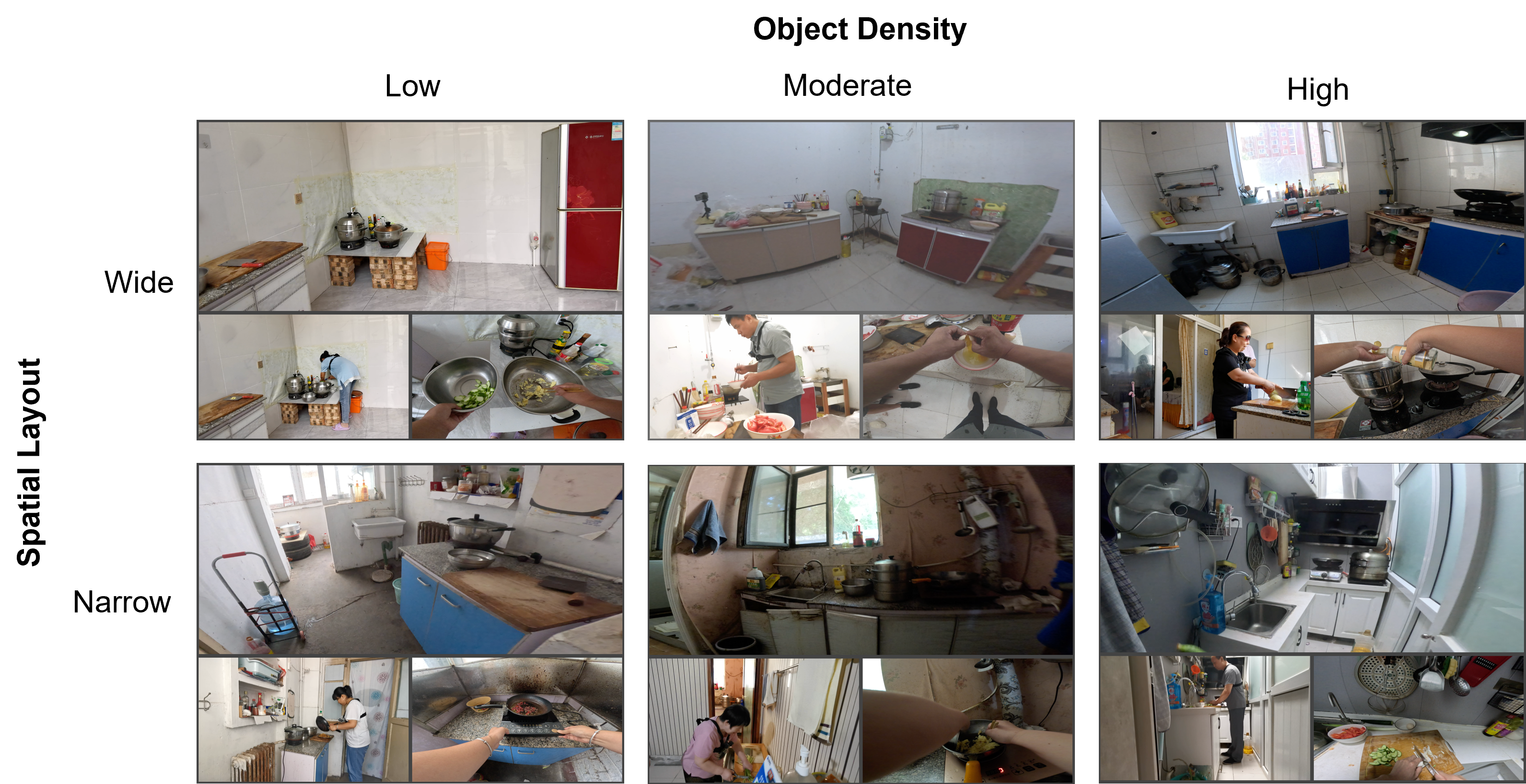}
  \caption{Contextual Inquiry: sample views of participants' home kitchens with varying spatial layouts (from wide to narrow), as well as object density.}
  \Description{There are six sets of images in this figure (2*3). Each set of image has a subfigure of an overall kitchen space that shows the kitchen layout, a subfigure of our participants making the dish that is captured through a stationary camera, and an egocentric view from the chest-worn camera to show the participant actually making the meal. Ot the left side of the figures, there are texts to indicate the top row is wide and the bottom is narrow for spatial layout. At the top side, it shows object density from left to right as low, moderate, and high.}
  \label{fig:teaser}
\end{figure*}

Assistive vision systems like SeeingAI \cite{SeeingAI86:online} and TapTapSee \cite{TapTapSe75:online} have proven effective in helping visually impaired people identify objects under various settings. However, these systems frequently fall short in the kitchen due to usability challenges and impractical designs \cite{li2021non}. For instance, past studies have highlighted their lackluster performance in recognizing clusters of objects or retrieving precise information about an object, such as the expiration date of products \cite{Howcanwe1:online,Microsof21:online}. Additionally, these systems often require holding a smartphone to capture information, which is inconvenient and impractical during cooking tasks that are time-sensitive \cite{li2021non}. Therefore, it is crucial to understand how visually impaired individuals obtain contextual information about kitchen objects as well as explore effective ways to communicate this information while cooking, thereby shedding light on opportunities to increase culinary independence and enhance their overall quality of life.

To summarize, in this work, we aim to investigate:

\begin{itemize}
    \item [RQ1] What contextual information about objects is important to people with vision impairments during cooking?
    \item [RQ2] What techniques and strategies do blind cooks use to non-visually gather and utilize information about objects during cooking and food preparation tasks?
    \item [RQ3] What are the most effective methods for conveying necessary contextual information about objects at the right times and in the right ways?
\end{itemize}

To address our research questions, we conducted a comprehensive contextual inquiry study with 12 visually impaired individuals experienced in cooking (Figure \ref{fig:teaser}). This inquiry took place in participants' own kitchens, where they were asked to naturally prepared meals using ingredients available in their refrigerators (as detailed in Section \ref{Contextual Inquiry}). Subsequently, they were instructed to perform specific kitchen-related tasks related to object recognition, such as identifying ingredients and organizing their kitchen space). In the end, we conducted semi-structured interviews to gain further insights into the motivations and rationales behind their actions, understand their needs for object-related information, and explore their views on integrating AI-powered assistive technologies in the kitchen (Section \ref{semi-structured interview}).

Our study offers a comprehensive analysis of the contextual information requirements for objects, comprising five primary, two secondary, and one application-specific category (Section \ref{Findings: Contextual Information Need of Objects (RQ1)}). We explore the nuanced process of gathering contextual information about objects for cooking activities, which involves establishing intentional contextual associations with objects (Section \ref{Findings: Process of Obtaining Contextual Information of Objects for Cooking Activities (RQ2)}). Furthermore, we examine factors related to displaying and communicating contextual information, including adjustable information verbosity (Section \ref{Findings: Contextual Information Presentation and Deployment Considerations (RQ3)}). Finally, to provide a holistic perspective, we discuss the design implications for AI-powered assistive technologies tailored for visually impaired individuals in kitchen environments (Section \ref{Discussion}).

Our work makes the following contributions:

\begin{itemize}
    \item A \textbf{contextual inquiry study} with 12 visually impaired individuals experienced in cooking, conducted in their own kitchens, yielding novel insights into their challenges and needs regarding objects' contextual information.
    \item A \textbf{taxonomy of objects' contextual information need} for cooking, including primary, secondary, and application-specific information, which fills a notable gap in existing research and offers a systematic guide for developing future systems to support people with vision impairments in the kitchen.
    \item A documentary of the \textbf{existing process of obtaining contextual information non-visually and associated challenges}, highlighting how visually impaired cooks create intentional associations with objects, enriching our understanding of their cooking experiences.
    \item A discussion of \textbf{the strategies for presenting and communicating contextual information} with future AI-powered assistive technologies.
\end{itemize}

\section{Related Work}
In this section, we first present the background knowledge of cooking experiences by people with vision impairments. We then describe the related work of information identification for people with vision impairments, followed by existing AI-powered technologies for kitchen space.

\subsection{Cooking Experiences by People with Vision Impairments}

Vision impairments have been shown to significantly affect individuals' experiences related to food, eating, and cooking \cite{bilyk2009food}. In a study involving over 100 visually impaired individuals, Jones et al. discovered a substantial correlation between the severity of vision impairment and the challenges faced in shopping for ingredients and preparing meals \cite{jones2019analysis}. This research also highlighted a concerning connection between vision impairments and malnourishment, ultimately leading to a diminished quality of life \cite{jones2019analysis}. Bilyk et al. \cite{bilyk2009food} conducted semi-structured interviews with nine visually impaired individuals, revealing their heavy reliance on prepared food from external sources. Notably, all participants in the study reported consuming a minimum of 40\% of their dinners in restaurants to avoid the challenges associated with cooking \cite{bilyk2009food}. More recent findings from Kostyra et al. \cite{kostyra2017food} indicated that out of 250 survey respondents, 49.6\% of visually impaired individuals prepare their meals independently, while others seek assistance from sighted and/or blind individuals. Notable difficulties in food preparation included peeling vegetables (82.1\% reported difficulty) and frying foods (72\% reported difficulty). Conversely, tasks not requiring heat or specialized tools, such as preparing sandwiches and washing fruits, were reported as more manageable. Given the ease of preparation, 57.6\% of participants opted for ready-to-eat products, while only 14\% preferred ready-to-heat meals \cite{kostyra2017food}.

Prior research has highlighted different practices and challenges that people with vision impairments have in cooking activities \cite{li2021non,bilyk2009food,wang2023practices}. For example, Li et al. \cite{li2021non} uncovered various difficulties for people with vision impairments while cooking, such as measuring, organizing space, tracking objects, and quality inspection. Among the tasks described by people with vision impairments in the kitchen \cite{li2021non,bilyk2009food,wang2023practices}, such as tracking objects and organizing space, many are relevant to the identification of different contextual information of objects (e.g., shape, color, location). Despite these existing explorations into the cooking experiences of people with vision impairments, there remain unclear regarding what contextual information of objects is needed across different cooking procedures.

\subsection{Information Identification for People with Vision Impairments}
Cooking often requires people with vision impairments to obtain contextual information about objects while cooking. Prior research has explored opportunities of adding tactile markers to devices or objects \cite{suzuki2017fluxmarker,guo2017facade}. For example, Guo et al, \cite{guo2017facade} created 3D printed tactile marking to better support people with vision impairments to interact with different interfaces. Beyond adding tactile markers, prior work also explored using crowdsourcing \cite{bigham2010vizwiz,gurari2018vizwiz} or computer vision \cite{guo2016vizlens,fusco2014using,morris2006clearspeech,tekin2011real,kianpisheh2019face,bigham2010vizwiz} to identify objects of interest. For example, Vizwiz \cite{bigham2010vizwiz} introduced a crowdsourcing-based approach for mobile phones that answers visual questions in nearly real-time, such as the color of objects. Moreover, VizLens leveraged computer vision and crowdsourcing to enable people with visual impairments to interact with different interfaces, such as a microwave oven \cite{guo2016vizlens}. Beyond supporting object recognition, Zhao et al. \cite{zhao2016cuesee} also explored how should visual information be presented to people with vision impairments and how should such system guide people to the targetted objects. Given existing approaches to identifying contextual information of objects by people with vision impairments, little has been explored to understand what are the contextual information needs in cooking scenarios, as well as how such systems should be developed to support cooking-related tasks.

\subsection{\red{AI-Powered Technology for Activities of Daily Living in the Kitchen}}
\red{AI-based kitchen technologies have been widely explored in HCI to enhance people's quality of life, such as monitoring kitchen activities and objects \cite{olivier2009ambient,lei2012fine}, supporting multimodal control and automation with kitchen appliances \cite{kim2017study,vu2018application,blasco2014smart}, enabling sensing capabilities for smart tools and utensils \cite{konig2015lab}, and providing dynamic guidance for cooking instructions \cite{ficocelli2012design,chen2010smart,kusu2017calculating,chang2018recipescape}. For example, Lei et al. \cite{lei2012fine} deployed an RGB-D camera inside the kitchen space and used RGB-D cameras to recognize fine-grained activities that include both activity and object recognition. Furthermore, Konig and Thongpull \cite{konig2015lab} invented Lab-on-Spoon, a 3D integrated multi-sensor spoon system for detecting food quality and safety, such as temperature, color, and pH value, to differentiate ingredients like fresh oil vs. used oil. Beyond recognizing activities and objects with specific technologies, prior research also stated the importance of providing full experiences for people in the kitchen space with deployments of both hardware and software \cite{blasco2014smart}. Although prior research has explored different AI-powered applications for kitchen activities it is unknown what contextual information is important to people with vision impairments in the kitchen and how systems such as these might need to be adapted for people with vision impairments.}

\section{Contextual Inquiry Study of Blind Cooking}
\red{Contextual inquiry has been used in field research to understand end users' experiences, preferences, and challenges during their everyday activities \cite{raven1996using,karen2017contextual,holtzblatt1995contextual,beyer1999contextual,liu2019unakite,liu2021reuse}. Contextual inquiry research can support deeper understandings of everyday human behaviors through surfacing facts, details, constraints, and structures \cite{beyer1999contextual}. Contextual inquiry has been widely adopted as a research method to better understand marginalized groups, cultural learning, and accessibility practices (e.g., \cite{de2004introducing,dosono2015m,macleod2017understanding,stangl2020person}).} To investigate our research inquiries comprehensively, we initiated a \textbf{contextual inquiry involving 12 participants with vision impairments naturally cooking in their kitchens}. Our study unfolds in three distinct phases: the pre-study interview, the contextual inquiry, and the semi-structured interview.

\subsection{Participants}
We recruited 12 people with vision impairments from the mailing list of the China Disabled Persons' Federation (Table \ref{table:participants}). To participate in our study, participants were required to be 18 years or older, legally or totally blind, and had prior experience with cooking. Among the 12 participants we recruited, six of them are female and six are male (Table \ref{table:participants}). The average age of our participants was 42.5 (SD = 6.1). Eight of them are totally blind and four are legally blind (Table \ref{table:participants}). \red{Regarding the four participants who are legally blind, P1 has no vision in the left eye and light perception in the right eye. P3, P8, and P12 have some light perception in both eyes.} They had an average of 21.7 years of cooking experience (SD = 9.0). As per their self-reports, seven of them cook every day, one cooks three or four times a week, two of them cook once per week, one cooks two or three times a month, and one cooks once per month (Table \ref{table:participants}). Regarding living arrangements, eight resided with their families, one with roommates, and the remainder lived independently. (Table \ref{table:participants}). Participants were compensated in local currency which is equivalent to \$40 USD. The recruitment and study procedure was approved by our organization's Institutional Review Board (IRB). Each participant's engagement took approximately 90 to 105 minutes.

\begin{table*}[t]
\begin{tabular}{l|lllll}
\toprule
\textbf{PID}  & \textbf{Age} & \textbf{Gender} & \textbf{Vision Condition} & \textbf{Cooking Experiences and Frequency} & \textbf{Living Condition} \\ \midrule\midrule
\multicolumn{1}{l|}{P1}  & 41           & Male            & Legally Blind             & 16 Years, Three or Four Times a Week       & Living with Family        \\\midrule
\multicolumn{1}{l|}{P2}  & 40           & Male            & Totally Blind             & 18 Years, Once per Month                   & Living with Family        \\\midrule
\multicolumn{1}{l|}{P3}  & 32           & Female          & Legally Blind             & 12 Years, Everyday                         & Living with Family        \\\midrule
\multicolumn{1}{l|}{P4}  & 46           & Male            & Totally Blind             & 24 Years, Once per Week                    & Living with Family        \\\midrule
\multicolumn{1}{l|}{P5}  & 52           & Male            & Totally Blind             & 29 Years, Once per Week                    & Living with Family        \\\midrule
\multicolumn{1}{l|}{P6}  & 44           & Female          & Totally Blind             & 15 Years, Everyday                         & Living with Family        \\\midrule
\multicolumn{1}{l|}{P7}  & 49           & Male            & Totally Blind             & 26 Years, Everyday                         & Living Independently      \\\midrule
\multicolumn{1}{l|}{P8}  & 36           & Female          & Legally Blind             & 10 Years, Everyday                         & Living with Family        \\\midrule
\multicolumn{1}{l|}{P9}  & 39           & Male            & Totally Blind             & 18 Years, Everyday                         & Living Independently      \\\midrule
\multicolumn{1}{l|}{P10} & 42           & Female          & Totally Blind             & 36 Years, Everyday                         & Living Independently      \\\midrule
\multicolumn{1}{l|}{P11} & 51           & Female          & Totally Blind             & 38 Years, Everyday                         & Living with Friends       \\\midrule
\multicolumn{1}{l|}{P12} & 38           & Female          & Legally Blind             & 18 Years, Two or Three Times a Month       & Living with Family        \\ \bottomrule
\end{tabular}
\caption{Demographic information of our study participants}
\label{table:participants}
\end{table*}

\begin{figure}[t]
    \centering
    \includegraphics[width=1\columnwidth]{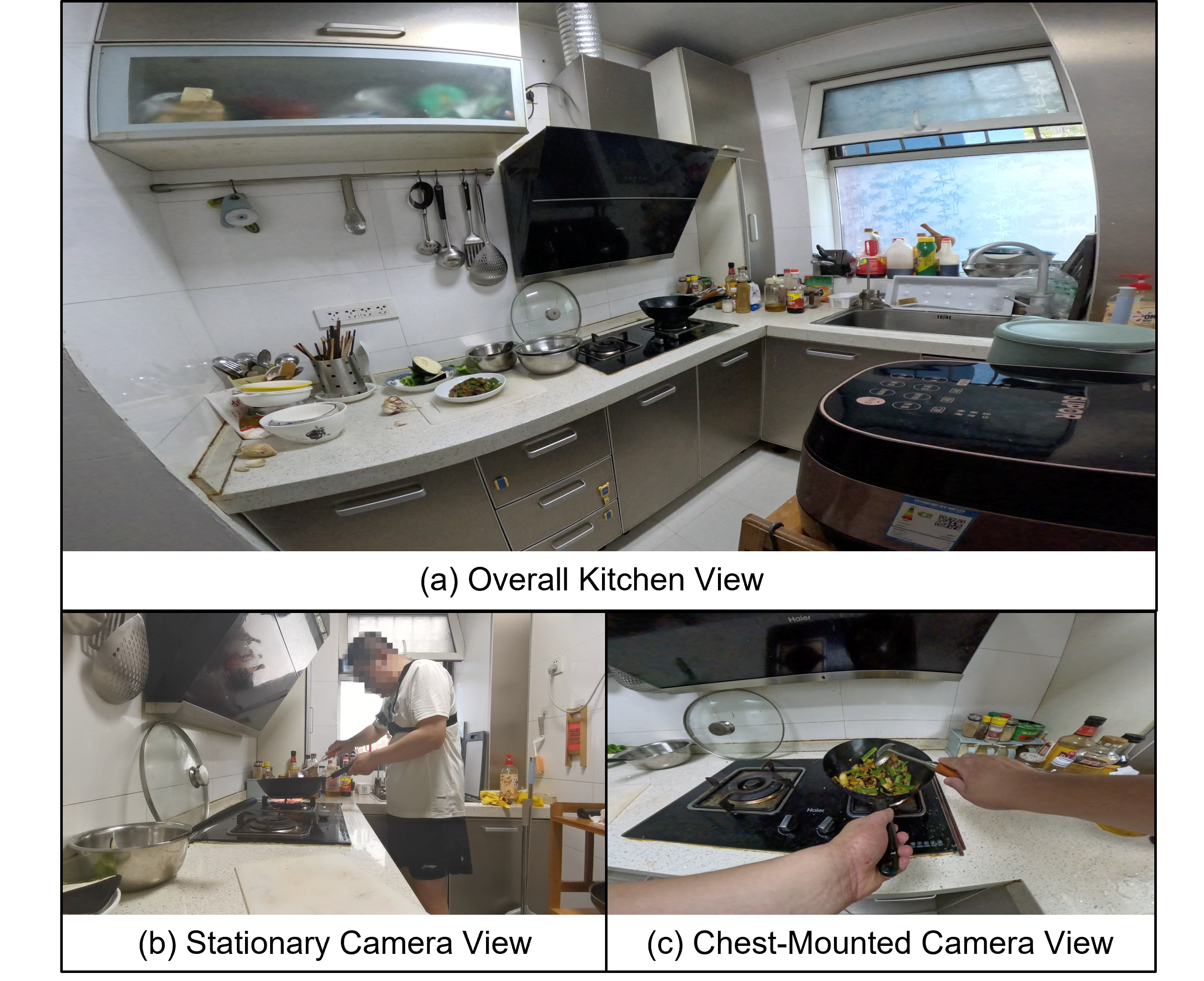}
    \caption{Contextual inquiry settings. Three camera views captured during the study to show the overall kitchen environment, ego-centric view, and a 3rd-person camera view of the person cooking. The top image is the overall still image of the kitchen setting. The bottom left image is the view from the stationary camera that shows our participant making a meal. The bottom right image is the egocentric view captured from the chest-mounted camera.}
    \label{fig:setting}
    \Description{This figure contains three subfigures that show the contextual inquiry settings. The top figure is the overall still image of the kitchen setting, there is a kitchen exhaust fan, a gas range, different utensils, a water sink, and several cabinets in the space. The bottom left figure is the view from the stationary camera that shows our participant making a meal. The bottom right figure is the egocentric view captured from the chest-mounted camera, it shows a person making a stir-fry.}
\end{figure}

\subsection{Study Procedure}
Our study unfolds in three distinct phases: a pre-study interview, a contextual inquiry, and a semi-structured interview.

\subsubsection{Pre-study Survey [5 Minutes]}
In the initial pre-study interview, we gathered demographic information from our participants. This included details such as age, gender, vision condition, cooking experience, cooking frequency, living arrangements, preferred cooking activities, and any challenges they encountered in the culinary domain.

\subsubsection{Contextual Inquiry [75 Minutes]}
\label{Contextual Inquiry}

\begin{figure}[t]
    \centering
    \includegraphics[width=1\columnwidth]{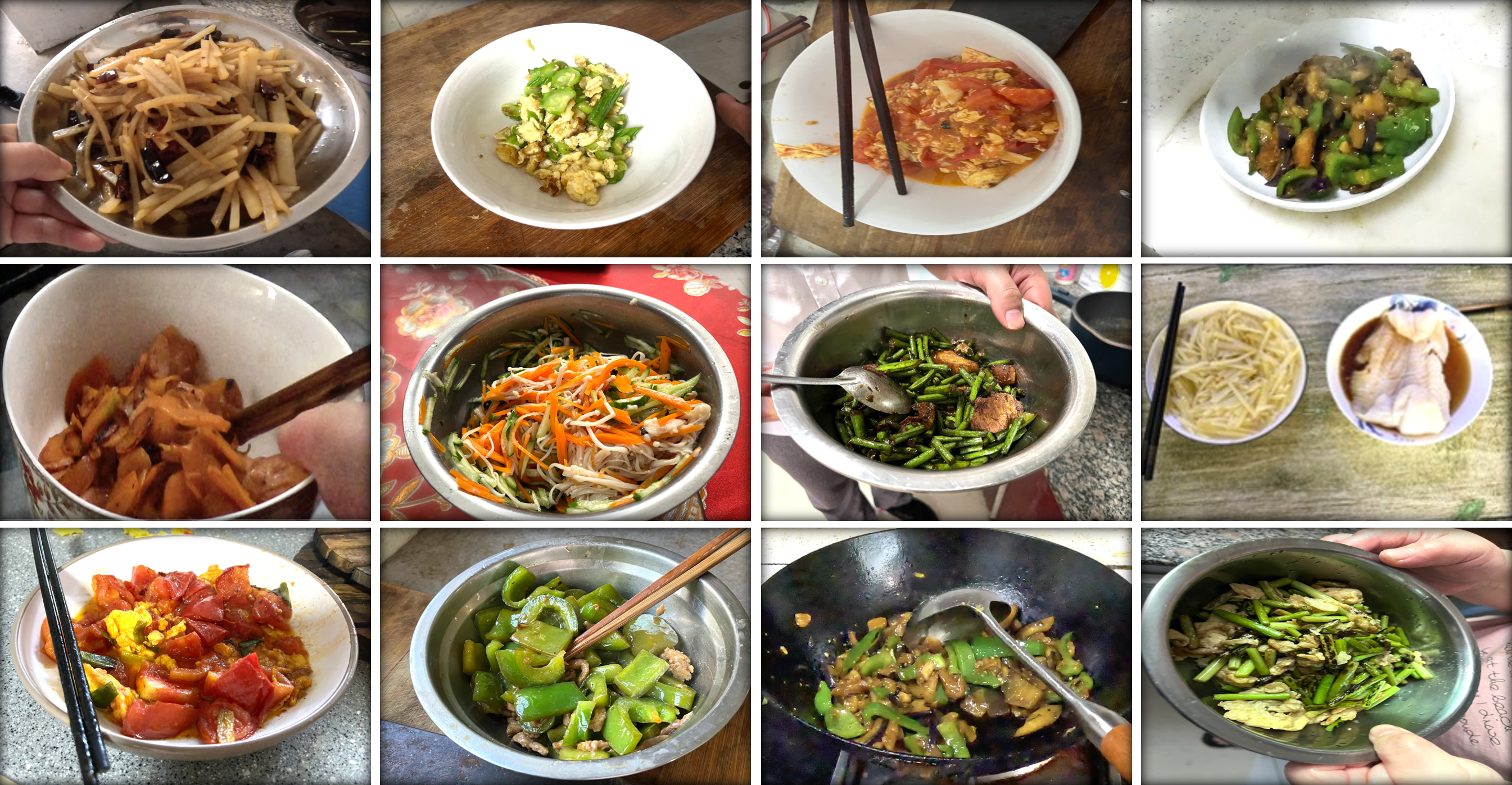}
    \caption{Dishes made by our participants during the contextual inquiry in their kitchens.}
    \label{fig:outcome}
    \Description{This figure contains 12 subfigures. Each subfigure represents a dish made by our participant. From left to right, top to bottom are, stir-fried sliced potato with chili pepper, stir-fried cucumber with egg, stir-fried tomato with egg, stir-fried eggplant with green pepper, stir-fried carrots, salad with enoki mushroom, carrot in vinegar, stir-fried green beans with pork, steamed fish in soy sauce, stir-fried tomato with egg, stir-fried green pepper with meat, stir-fried potato with green pepper, stir-fried garlic sprout with egg.}
\end{figure}

During the contextual inquiry phase, participants engaged in various cooking activities within their personal kitchens (as depicted in Figure \ref{fig:setting}). To record their culinary experiences comprehensively, participants were equipped with a GoPro 11 camera \cite{GoProHER29:online} attached to their chests, allowing us to document their actions and behaviors (as shown in Figure \ref{fig:setting}). To provide a comprehensive view of their activities, we installed a stationary camera within their kitchen environments (Figure \ref{fig:setting}). In the contextual inquiry, there were two main tasks: 

\red{\textbf{[TASK 1: Self-Directed Cooking]}: To observe a full experience of cooking, our participants were asked to first explore the food and ingredients that they have in their own kitchen, and then make a dish based on the availability of ingredients} (See Figure \ref{fig:outcome} for dishes made by our participants). After making the dish, participants were asked to serve the dish and clean the kitchen. During the cooking process, participants were encouraged to vocalize their thoughts while performing tasks, following the think-aloud protocols throughout the contextual inquiry \cite{van1994think}.
    
\textbf{[TASK 2: Specific Cooking-related Activities]}: Following the completion of \red{Task 1}, participants were further engaged in specific cooking-related activities to gain a deeper insight into their procedural methods and information requirements. Drawing from prior research findings \cite{li2021non}, we chose eight key activities that epitomize the essential cooking tasks for individuals with vision impairments that are relevant to object identification. These activities encompassed: 1) identification of ingredients and food items, 2) recognition of cookware and utensils, 3) precise measurement, 4) monitoring of cooking progress, 5) the process of serving food, 6) ensuring safety measures, 7) maintaining kitchen organization, and 8) executing grocery shopping. Participants were requested to demonstrate their typical approach to completing these tasks and to vocalize their thought processes throughout, adhering to the think-aloud protocols \cite{van1994think}.


\subsubsection{Semi-structured Interview [25 Minutes]}

\label{semi-structured interview}

Following the completion of the contextual inquiry, we proceeded to conduct a structured semi-structured interview. This interview served as a platform for a comprehensive debriefing, allowing us to delve into our observations and findings from the contextual inquiry. Additionally, it provided an opportunity to explore various facets of future design considerations related to the communication of visual information and the form factors of assistive technologies within the kitchen environment.

\textbf{[Debriefing of the Contextual Inquiry]}: In this segment of the interview, we initiated discussions concerning the behaviors and processes observed during the contextual inquiry (refer to Section \ref{Contextual Inquiry}). We delved deeper into the challenges encountered by participants, as well as their specific visual information needs. This encompassed topics such as object localization, contextual information regarding objects, and the dynamics of visual information requirements throughout different stages of the cooking process.

\textbf{[Design of Information Communication and Deployment]}: Subsequently, our conversation shifted towards the design aspects related to information communication and deployment within the kitchen setting. We inquired about participants' preferences for how information should be conveyed to them while cooking, their information needs during culinary activities, their opinions on the form factors of assistive technologies within the kitchen, and any other concerns or considerations they wished to share.



\subsection{Data Analysis}
The contextual inquiries were meticulously documented using both video recording methods. High-resolution footage was captured at 5.3k resolution with a frame rate of 30Hz utilizing the GoPro 11 camera in HyperView \cite{GoProHER29:online}. Additionally, a stationary camera was employed to supplement the recordings. Meanwhile, the semi-structured interviews were recorded in audio format. \red{We leveraged the video resources captured by the body-worn camera and the stationary camera for the \textbf{contextual inquiry} analysis \cite{knoblauch2012video}. For the analysis of the contextual inquiry videos, we used thematic analysis \cite{braun2006using}. Two researchers independently annotated the video and open-coded the observations. Our analysis} centered on two key aspects: the contextual information needs related to objects and the processes involved in acquiring contextual information about these objects (as detailed in Section \ref{Contextual Inquiry}). As for the analysis of the \textbf{semi-structured interviews}, a similar thematic analysis approach was employed \cite{braun2006using}. This analysis revolved around themes related to the processes of acquiring contextual information, expected methods of information communication for objects, visual information requirements, and considerations regarding form factors (refer to Section \ref{semi-structured interview}).





\section{Findings: Contextual Information Needs of Objects (RQ1)}
\label{Findings: Contextual Information Need of Objects (RQ1)}
In this section, we first present the five fundamental contextual information needs of objects that people with vision impairments prefer while cooking: position information, orientation information, proximity and grouping information, similarity and duplicate information, and internal state information. We further present three secondary and application-specific information needs that are relevant to kitchen activities: safety-related information, health-related information, and plating and serving information. We illustrated this in the Table \ref{table:contextual}.

\begin{table*}[]
\begin{tabular}{p{3.6cm}|p{8cm}|p{2cm}}
\toprule
\textbf{Attribute}                  & \textbf{Description}                                                                                                                                                                                              & \textbf{Priority}             \\ \midrule
Position                   & The relative location to a reference point, or an ``anchor,'' to indicate the object's position                                                                                                          & Primary              \\ \midrule
Orientation                & Information about how an object is currently oriented (e.g., vertical, horizontal, or tilted) relative to a reference point (e.g., human, object)                                                        & Primary              \\ \midrule
Proximity and Grouping     & Information about groupings of objects relative to others and the environment                                                                                                                            & Primary              \\ \midrule
Similarity and Duplicates  & Information about similar or duplicate objects, which includes differentiation between similar objects, relative positions between objects that are similar, and the overall quantity of similar objects & Primary              \\ \midrule
Internal State             & Information about internal state of objects, such as cleanness, freshness, boiling water, the amount of solid or liquid inside the container, and the doneness of food                                   & Primary              \\ \midrule
Safety-related Information & Information to monitor anything that might be harmful (e.g., knock over objects, flamming)                                                                                                               & Secondary            \\ \midrule
Health-related Information & Information to track objects with health hazards after consumption (e.g., expired food, overcooked food)                                                                                                 & Secondary            \\ \midrule
Plating and Serving        & Information about the final appearance and presentation of objects upon finishing (e.g., color distribution)                                                                                             & Application-Specific \\ \bottomrule
\end{tabular}
\vspace{1mm}
\caption{\red{Contextual information needs revealed during our study. Primary attributes were required to interact with objects, while secondary and application-specific attributes were related to understanding the state of objects and manipulating them.}}
\label{table:contextual}
\end{table*}

\subsection{Five Fundamental Categories of Contextual Information for Objects}

\subsubsection{Position Information}
\label{location information}
During our contextual inquiry conducted in the participants' kitchens, we uncovered a unanimous consensus among our participants regarding the paramount importance of knowing the precise position of objects within their culinary domains, which could reduce the time effort of finding objects while cooking (9), and support autonomy and agency in the kitchen (3). \red{We found over half of our participants encountered difficulties when attempting to locate specific items while cooking, due to the mental load of memorizing the positions (P1, P8, P11) or misplacement of items by their family members or friends (P2, P4, P6, P10). This challenge increased when dealing with less frequently used items. To remember object locations, our participants typically advocated for the \textbf{use of a reference point in the kitchen, or an ``anchor,'' to indicate the object's  position}}. For instance, they preferred describing the oyster sauce as being "on the windowsill" rather than specifying its coordinates in 3D space (e.g., x, y, z). P5 elucidated:

\begin{quote}
    ``To determine the location of the item I'm searching for, I simply require a relative position in relation to a reference point in my kitchen, such as near my gas range or on my fridge.''
\end{quote}

\subsubsection{Orientation Information}
Our inquiry also shed light on another critical facet - the need to ascertain the orientation of various kitchen objects, signifying their \textbf{current alignment or positioning relative to a specific reference point or another object}. We found that our participants encountered challenges in discerning the orientation of objects during different culinary tasks (P2, P4, P5, P6, P9, P11). For example, both P2 and P5 grappled with slicing pork belly with the correct orientation, often leading to the unintended separation of fat and lean meat portions. Similarly, our participants faced recurrent tribulations when endeavoring to align a wok precisely with plates or bowls during the serving process (P4, P6, P9, P11). This issue invariably resulted in unwanted food spillage. P11 provided illuminating insights into the complexities of this matter:

\begin{quote}
    ``Perfectly aligning the wok with the plate during serving can be quite challenging, and as a consequence, I frequently find myself grappling with food spillage, necessitating subsequent cleanup efforts.''
\end{quote}

\subsubsection{Proximity and Grouping Information}
\label{spatial information}
In addition to object orientation, our participants placed significant emphasis on contextual information related to proximity and grouping - essentially, the \textbf{arrangement and organization of objects in relation to one another and the kitchen environment}. Knowing the proximity and grouping information of certain object groups can support maintaining the kitchen, as well as navigating the kitchen space. This aspect encompassed the pressing need to determine if objects were correctly placed, if any misplacements or disarray had occurred (P4, P11), and whether alterations or rearrangements to the kitchen space had transpired during or after culinary activities (P5, P9). For instance, we observed that P5 faced difficulties locating sauces and ingredients following his daughter's cooking session, as she had inadvertently rearranged various items. P5 articulated his predicament:

\begin{quote}
    ``I usually...have my sauces like soy sauce and vinegar arranged on the second shelf and other solid ingredients such as sugar and salt carefully positioned on the third shelf. However, my daughter cooked a meal yesterday, and the displacement of items made it exceedingly challenging to locate things today.''
\end{quote}

\subsubsection{Similarity and Duplicates Information}
Our research also illuminated the critical need to acquire \textbf{information about similar or duplicate objects}, encompassing differentiation between similar items, determining the relative positions of objects, and assessing the overall quantity of similar objects. Understanding the quantity of specific items, such as tomatoes within the fridge, emerges as vital for monitoring food supplies, particularly when preparing for grocery shopping. During our study, we observed that P3 had multiple tomatoes located at various spots within her refrigerator. When asked about the tomatoes' whereabouts, she discovered two with holes and remarked:

\begin{quote}
    ``I wasn't aware that I had these tomatoes tucked away in the corner of my fridge, and I can't recall how long they've been there. It's possible my son placed them there.''
\end{quote}

Moreover, we noted that providing information regarding the differentiation between similar objects can significantly assist individuals with vision impairments in comprehending their kitchen environments. For example, P1 and P8 both expressed the desire to know how different plates were stacked together on a shelf and whether all the plates were identical or if any variations existed during the serving process.

\subsubsection{Internal State Information}
\label{status of objects}

\red{Kitchen objects may possess various \textbf{internal states, including temperature, freshness, cleanliness, the condition of solid or liquid contents, and the degree of doneness for food that is being cooked}. Our participants noted that ascertaining these internal states can be a challenge, primarily because many of these assessments rely on visual cues. For instance, our participants expressed difficulties in determining the cleanliness of vegetables or meat (P2, P9). Other challenging tasks include monitoring the water temperature to determine if it has reached boiling point (P5) or gauging the readiness of food (P6). P9 explained this issue:}

\begin{quote}
    ``It's impossible for me to determine if vegetables are clean or not during the washing process. Consequently, I often find myself repeatedly washing them to ensure their absolute cleanliness.''
\end{quote}


\subsection{Secondary and Application-specific Information}
In addition to the five fundamental contextual information categories, we identified three types of information that people with vision impairments are acutely aware of while interacting with objects in the kitchen.

\subsubsection{Safety-related Information}
\label{safety-related information}
The first secondary information is safety-related information, which encompasses monitoring objects that have the potential to cause harm. This includes objects that could be accidentally knocked over during kitchen tasks (P4, P9, P10) and safety-related concerns, such as monitoring the temperature of cooking equipment to prevent accidents (P4, P5, P6, P10, P11).
Our observations revealed instances where P4 accidentally knocked over a salt bottle while searching for a plate, and P9 tipped over a nearby water cup. P9 emphasized the significance of being aware of potential obstacles:

\begin{quote}
    ``Being aware of potential obstructions that I might knock over could greatly benefit me throughout the preparation and cooking process. It could also help reduce the anxiety associated with interacting with objects while learning.''
\end{quote}

\subsubsection{Health-related Information}
\label{health-related information}

Our contextual inquiry highlighted the critical importance of knowing contextual information about objects for health-related considerations. This encompasses factors like checking expiration dates (P1, P10, P12), identifying food with potential health risks, such as overcooked items (P2, P5), and recognizing visual cues on food items, such as stickers on vegetables (P6, P8). P2 articulated the challenges he faced in this regard:

\begin{quote}
    ``To ensure all of the meat is fully cooked, I usually cook it for a longer period, which sometimes had some of the food got overcooked or even burned.''
\end{quote}

\begin{figure}[t]
    \centering
    \includegraphics[width=1\columnwidth]{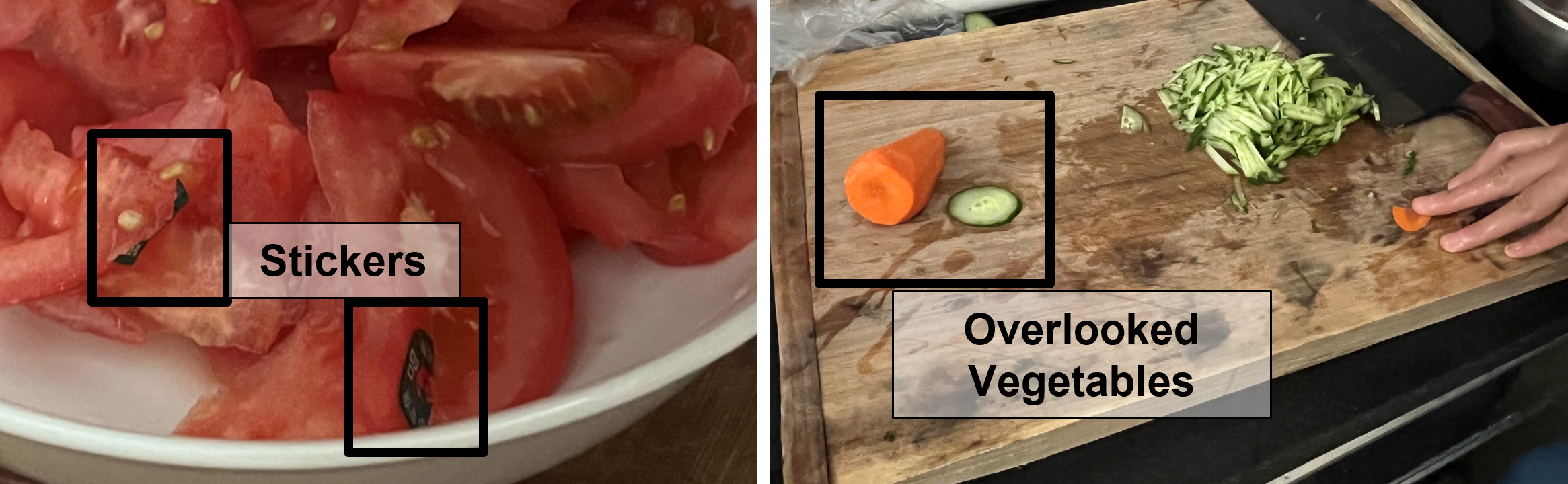}
    \caption{\red{Some visual information can have health implications. Left: Sliced tomatoes still have stickers attached (highlighted); Right: Some vegetables were overlooked and left on the cutting board (highlighted).}}
    \label{fig:health}
    \Description{Two figures. The left figure shows cut tomatoes and some pieces of them have stickers on them. There are two black boxes highlighting the stickers with the text ``Stickers'' on the side. The right figure shows a cut cucumber on the right-hand side of the cutting board. There is a remaining carrot and a slice of cucumber at the left side of the cutting board that is highlighted with a black box with the text ``Overlooked Vegetables''.}
\end{figure}

We also observed instances where participants unintentionally left behind or missed certain food items, especially when dealing with round-shaped objects like green beans, or during the transfer of items from the cooking vessel to the plate. This oversight could lead to health-related concerns, such as consuming spoiled food or creating conditions favorable to pests like cockroaches (P2, P4, P5, P6, P11) (Figure \ref{fig:health}). For example, P5 inadvertently included a food sticker while slicing a tomato, and it was subsequently cooked in the dish (Figure \ref{fig:health}). P5 expressed:

\begin{quote}
    ``I had no idea there were stickers on the tomato, and it's challenging to use my hands to feel the entire tomato to detect the sticker. Wouldn't it be better if they stopped using stickers altogether?''
\end{quote}

\subsubsection{Plating and Serving Information}
Upon completing the cooking process, we uncovered our participants' preferences for being informed about the presentation and final appearance of the dishes they prepared. This encompassed details such as the arrangement of items on the plate (position), the degree of cooking (internal state), the spatial relationship with other items on the plate (proximity), and the orientation of food items (orientation). P5 emphasized the value of being aware of the visual presentation of their culinary creations:

\begin{quote}
    ``It's valuable for me to be aware of how appealing my food looks once it's served, or if I should consider adding additional vegetables or meat to enhance the overall visual presentation.''
\end{quote}

\section{Findings: Process of Obtaining Contextual Information of Objects for Cooking Activities (RQ2)}
\label{Findings: Process of Obtaining Contextual Information of Objects for Cooking Activities (RQ2)}
In this section, we begin by discussing how individuals with vision impairments rely on multiple sensory inputs to acquire various types of contextual information [indicated in brackets] concerning objects during cooking activities. These sensory inputs encompass touch, sound, and smell. Subsequently, we explore two distinct approaches employed by our participants to streamline the process of identifying objects of interest: the creation of supplementary contextual information and the simplification of the contextual information acquisition process related to objects.

\begin{figure}[t]
    \centering
    \includegraphics[width=0.8\columnwidth]{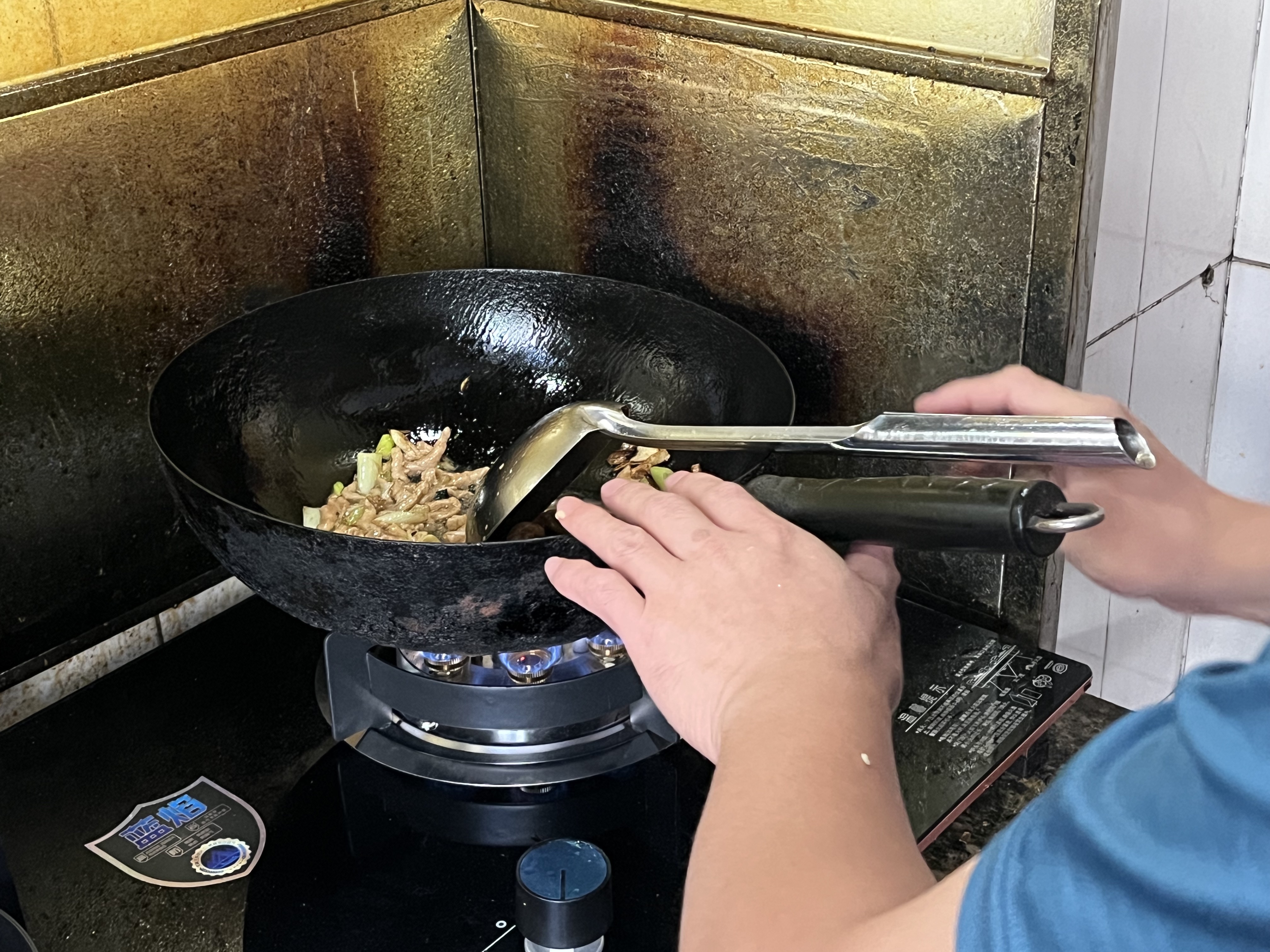}
    \caption{P9 using his left hand to touch the side of the wok to gauge the temperature of the wok.}
    \label{fig:touchwok}
    \Description{This figure shows a gas range and has a wok on top of it. There is a person who is using the right hand to spin the food by using a spatula and another hand to touch the side of the wok.}
\end{figure}

\subsection{\red{Multi-sensory Approaches to Obtaining Contextual Information}}
\label{multisensory}

\subsubsection{Touch}
\label{touch}
\red{Our contextual inquiry revealed that our participants relied on their sense of touch for tracking the status of objects in their kitchen, including dual-purpose scanning and memorization, manipulating objects, and performing safety inspections.}

\red{\textbf{Dual-purpose Scanning and Memorization:}
We found that our participants used touch for multiple purposes, such as localizing objects and checking the internal state of objects.} This included locating items like vegetables, fruits, and meat within the refrigerator, as well as identifying sauces, ingredients, or containers within the kitchen \textbf{[Position]}. \red{Touching food also communicated information about the} freshness of the food \textbf{[Internal State]}. P4 commented: \textit{``Using touch is my primary way of finding things in the kitchen, while I explore objects, I also feel the object to know if it is fresh through the stiffness or if there are holes on the skin.''}

During this process, they engaged in pre-organizing objects of interest and often memorized the positions of other similar or nearby objects as they scanned the space \textbf{[Proximity and Grouping][Similarity and Duplicates]}. For instance, P1 searched for eggplants in the fridge, scanning through it while simultaneously committing the location of garlic to memory. Later, when he needed minced garlic, he easily found it, remarking, \textit{``I memorized the position of the garlic last time when I was scanning through the fridge!''} However, we found this approach sometimes can take a long time to scan through objects, and people might miss certain objects through scanning due to the complexity of the space and form a wrong memory of objects and space (P3, P7, P11). P3 explained:

\begin{quote}
    ``It often takes me a while to find the vegetable that I want to get. And it is easy for me to miss some of it, because kitchen shelves and refrigerator storage are complex, such as my tomatoes were placed at multiple positions in the kitchen. Once I did not find it, then it might just stay in a corner for many days.''
\end{quote}

\textbf{\red{Precision and Manipulation:}}
While manipulating and interacting with objects, we found our participants also leverage touch to ensure objects are organized and aligned, \red{which maintained order and reduced the risk of spillage,} \textbf{[Orientation]} \red{as well as to count and measure} objects (e.g., sugar, vinegar) \textbf{[Internal State]}.
\red{To align objects, or when transferring materials between containers (e.g., adding sauces to salad, serving food from a wok) our participants usually used one hand to hold the object then used another hand to find and secure the other object}. To determine the quantity of dry ingredients such as salt and sugar, they relied on touch, using their hands to feel and specify the exact amount (P1, P4, P5, P8, P9). For liquids, they often placed a finger beneath the lid, allowing them to feel the liquid passing through their finger to gauge the quantity (P5, P7, P10). 

\textbf{Safety Inspection:}
Furthermore, we discovered that access to safety-related information could significantly reduce the risks associated with cooking for individuals with vision impairments. For example, both P9 and P10 routinely performed thorough inspections of flammable objects that are close to the gas range before cooking \textbf{[Proximity and Grouping][Safety]}. P10 expressed a desire for pre-cooking safety checks: \textit{``I would appreciate having some form of support for conducting safety checks before cooking to ensure there are no objects in close proximity to the range during cooking.''}
Additionally, we found that gauging the temperature of the wok \textbf{[Internal State][Safety]} presented one of the most formidable challenges for individuals with vision impairments, as this information was traditionally obtained through tactile means, such as direct touch (see Figure \ref{fig:touchwok}). This practice, although effective, often resulted in burns and blisters, as expressed by P9:

\begin{quote}
    ``I use my hand to feel the temperature of the wok; you can see my arm has many blisters and burns, but I have to use this method as there is no other way for me to gauge the temperature or balance the wok correctly.''
\end{quote}

\subsubsection{Sound}
\label{sound reliance}
\red{Recognizing and tracking sounds played a pivotal role in our participants' ability to assess the status of objects \textbf{[Internal State]} in cooking, such as temperature. As an illustrative example, P11 described a method involving the addition of a small amount of egg yolk to hot oil in a wok to listen to the resulting sound to estimate the oil's temperature. While sounds were usually helpful, it was sometimes difficult to follow sounds because of excessive background noise (9), such as kitchen exhaust fans or conversations with others in the room. P11 provided further insights into this issue:}

\begin{quote}
    ``I prefer using sound-based assessments, such as checking if water has boiled, but sometimes the differences in sound characteristics can be quite subtle...which is difficult to specify with my exhaust fan on.''
\end{quote}

In addition to assessing object status, participants utilized sound to estimate the quantity or volume of objects \textbf{[Internal State]}. An interesting example involved the use of containers with narrow nozzles. When pouring liquids into a pan or wok, the air inside the container compresses as the liquid flows, generating a distinct sound, often described as a ``burp.'' This auditory cue allowed individuals to approximate the amount of liquid dispensed. P8 detailed the practicality of this method for estimating quantities through sound while acknowledging its inherent limitations, such as reduced precision, particularly when the liquid level in the container was low: \textit{``I use sound to estimate how much oil I've poured into the pan. It's challenging to discern through touch alone. The first drop of oil hitting the wok generates a small sound, and since the oil bottle I purchased has only one nozzle, it doesn't continuously pour. Instead, it dispenses intermittently, producing a `burp' sound. I rely on this auditory feedback to gauge the amount of oil in the pan. However, it's not always precise, as the sound may become less noticeable when the liquid level in the bottle is low.''}

\begin{figure}[t]
    \centering
    \includegraphics[width=1\columnwidth]{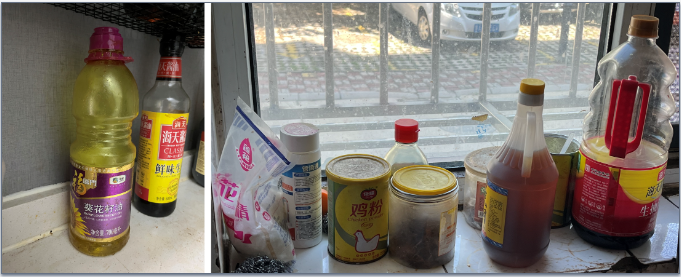}
    \caption{Unique containers differentiate contents. Participants use and reuse unique containers to help identify objects.} 
    \label{fig:refill}
    \Description{Two figures. The left one shows an oil bottle and a soy sauce bottle where the oil bottle has some oil stains. The right figure shows various bottles and containers with different shapes on a windowsill.}
\end{figure}

\subsubsection{Smell}

The sense of smell plays a critical role in helping individuals with vision impairments determine the internal state of objects and health-related information \textbf{[Internal State][Health]}, such as the freshness or the doneness of the food. In cases where touch exploration was not feasible or practical, participants relied on their olfactory senses to detect signs of spoilage. P3 illustrated this practice, stating, \textit{``It is common for us to have leftovers of main dishes as well as rice and bread. Sometimes, touching is often not feasible to assess the condition of the food. So, I use my sense of smell to determine if the food has gone bad. Spoiled food often emits a sharp and unpleasant smell due to fermentation.''} This reliance on smell allowed them to make informed decisions about whether it was safe to consume leftover food items.

Participants also utilized their sense of smell as a means to determine the readiness of certain dishes \textbf{[Internal State]}. Specific foods emitted distinctive aromas when they were close to being fully cooked. For example, P4 and P9 mentioned that particular dishes, such as those containing green peppers and meat, would release a savory aroma, signaling that they were nearly done. This olfactory cue served as an indicator of cooking progress. However, participants emphasized the importance of swift action once these aromatic cues were detected, as there was little room for delay between sensing the enticing aroma and preventing the food from becoming overcooked or burned. As P4 humorously put it, \textit{``I like to use smell to gauge the readiness of my food. When you catch that aroma, it feels like a culinary achievement. But don't celebrate too long; you need to promptly remove the food from the wok to prevent it from burning.''}

\subsection{\red{Altering Objects to Ease Identification}}
\label{customized creation of contextual information}
In addition to their multi-sensory strategies, our participants demonstrated the approach of actively creating additional contextual information for objects, thereby facilitating their recognition and organization. Through contextual inquiry, we uncovered a pervasive practice among all participants: the deliberate customization of objects to imbue them with supplementary contextual information. This approach involved introducing distinctive attributes, such as unique container shapes or deliberate organizational strategies, to enhance object identification and ensure secure organization within the kitchen.



\subsubsection{\red{Using Containers with Unique Shapes}}
We found that all of our participants typically chose to use different sizes or shapes of containers to indicate the difference between sauces, oil, or seasonings \textbf{[Similarity and Duplicates]} (Figure \ref{fig:refill}), which correspond to prior research showed that bartenders used different glasses to remember orders \cite{erickson2007hci}. From the observation of the contextual inquiry, we found that they easily spotted the sauce that they wanted, and we found that for P1, he used a thin-headed, mid-sized glass to store oyster sauce, a large, rectangle bottle to store vinegar, and a wide-headed, large plastic bottle to store soy sauce. P5 also refills oils to the bottle that he purchased a long time ago (Figure \ref{fig:refill}). P6 further explained:

\begin{quote}
    ``I personally use different shapes of containers to indicate the difference of oyster sauce, vinegar, and soy sauce. There was one time that my daughter purchased a new bottle of oyster sauce, which confused me for many days.''
\end{quote}



\subsubsection{Securing Objects for Organization}
\begin{figure}[t]
    \centering
    \includegraphics[width=0.8\columnwidth]{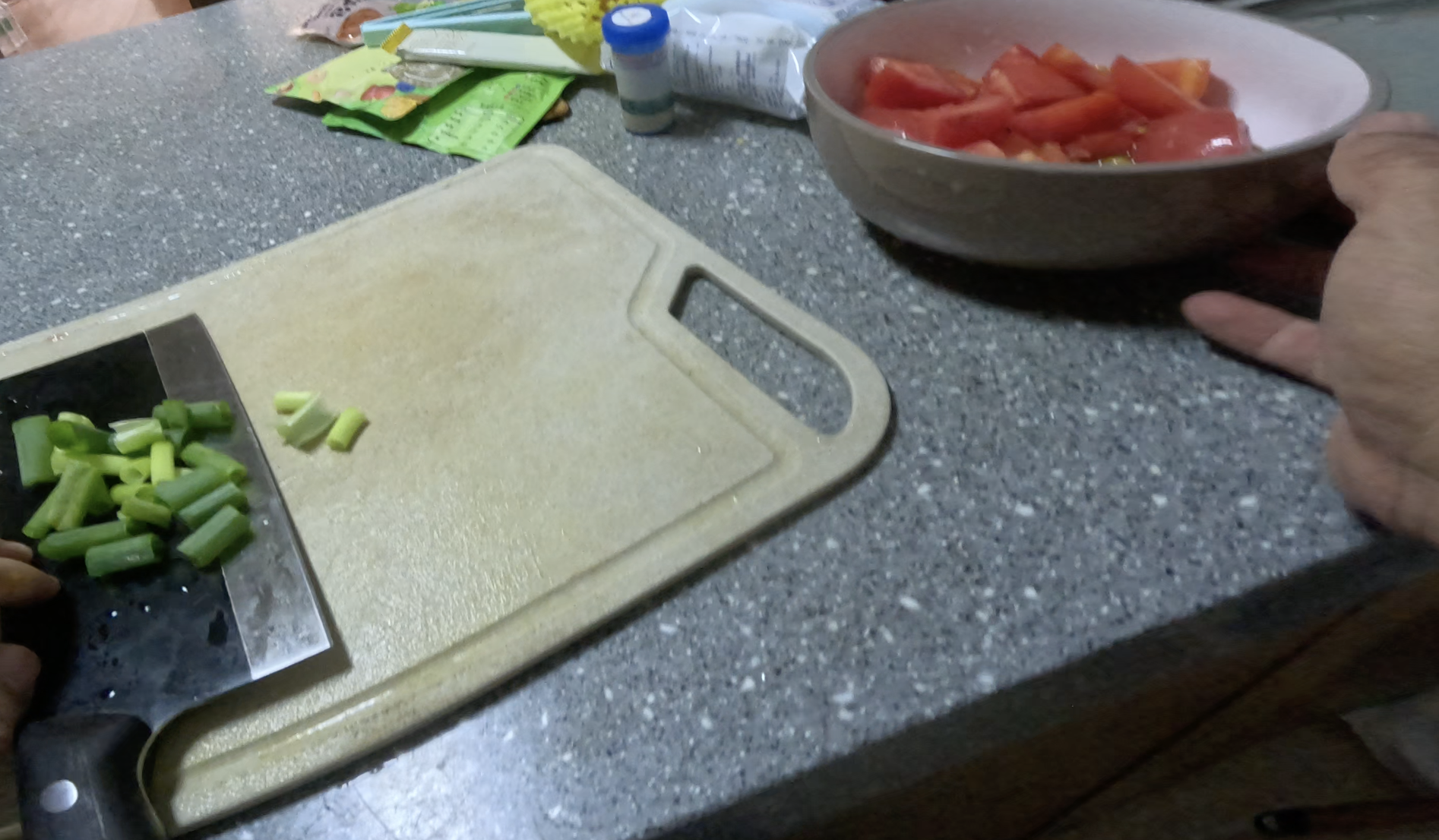}
    \caption{P4 used a knife as a ``container'' for scallions or minced garlic.}
    \label{fig:knife}
    \Description{There is a knife on a cutting board, and there are a good amount of cut scallions on the side of the knife.}
\end{figure}

Participants favored using bowls or plates to secure objects, promoting better organization, \red{rather than letting them roll freely on the counter} \textbf{[Orientation][Position]}. For example, P4, P5, P11, and P12 placed cut vegetables inside a bowl for ease of management. Similarly, P4, P6, and P10 employed a knife as a makeshift "container" to secure objects before cooking, such as scallions or minced garlic (Figure \ref{fig:knife}). This additional layer of context minimized the risk of accidents and lightened the cognitive load associated with memorizing object locations.

\subsubsection{Spatial Separation and Grouping of Objects}
In addition to enhancing the identification of individual objects, our study revealed that participants deliberately arranged objects in separate spatial groupings based on their purposes \textbf{[proximity and grouping]}, aiming to facilitate easier identification and access (P2, P3, P6, P8, P9, P11). During interviews, participants elaborated on the advantages of spatially separating objects, highlighting how it added another layer of context and improved the spatial recognition of object groupings. P8 provided insights into this behavior:

\begin{quote}
    ``I prefer to keep my kitchen essentials minimal and uncomplicated. Simultaneously, I adopt a practice of spatially separating different objects, aiding me in distinguishing between various items more effectively and reducing the chances of encountering unwanted objects due to spatial constraints.''
\end{quote}

Furthermore, participants mentioned their practice of organizing objects within dedicated mini-spaces, each containing a specific group of items. For instance, P11 stored all sauces in the same location, while P9 demonstrated how he arranged daily-use bowls within easily accessible spaces and placed less frequently used ones in cabinets:

\begin{quote}
    ``Locating and reaching certain spaces can be quite demanding, especially when I have to stoop down to access bowls near ground level. Therefore, I keep utensils I use daily on the counter, within easy reach, and store the others in cabinets.''
\end{quote}

\subsection{Optimizing Cooking Procedures}
\label{optimizing cooking procedures}
In their pursuit of greater convenience and efficiency, our participants devised a range of strategies to simplify their cooking procedures, reducing the need for extensive contextual information gathering. These innovative methods encompassed pre-assigning orders, sequential organization, strategic seasoning placement according to usage frequency, and preserving spatial arrangements during the cleaning process. Through these tactical approaches, they were able to optimize their cooking routines while minimizing the effort required to access contextual information about objects.

\subsubsection{Pre-assigned Orders and Sequential Organization}
Our study revealed that participants frequently prepared objects in a \textbf{specific order based on their sequential requirements} (P2, P5) \textbf{[Position]}. This approach involved completing preparation tasks before embarking on the actual cooking process, thus optimizing their time and effort. P2 exemplified this by arranging the bowl of oil in front of the onions, followed by the meat, to denote the order of preparing the dish: \textit{``Pre-arranging all the containers would save me time and effort during execution by minimizing the need to recheck the objects in the bowl.''} Furthermore, beyond arranging ingredients in the order they would be used, participants also \textbf{organized seasonings according to their frequency} of use (P9, P11) \textbf{[Similarity and Duplicates]}. P9 elaborated:

\begin{quote}
    ``I typically position the seasonings I use daily closer to the gas range, while placing others in cabinets or higher shelves.''
\end{quote}

Additionally, participants maintained the \textbf{spatial arrangement of objects} even during the post-cooking cleaning process (P4, P5, P7, P12) \textbf{[Proximity and Grouping]}. This practice served a dual purpose, as explained by P5:

\begin{quote}
    ``The cleaning process not only ensures cleanliness but also guarantees that all objects are returned to their designated positions, preventing difficulties in locating ingredients during subsequent cooking sessions.''
\end{quote}

\subsubsection{Step and Purpose Combination}
In pursuit of greater cooking efficiency, our participants skillfully combined specific cooking steps and harnessed objects for multiple purposes, simplifying their culinary endeavors, which reduced the need to obtain contextual information of multiple positions as well as differentiating different objects \textbf{[Position][Similarity and Duplicates]}. These strategies included merging cooking steps and employing objects with dual functionalities. Participants deliberately \textbf{integrated particular cooking steps and repurposed objects, effectively streamlining their culinary processes}. For instance, P3 favored using a pot as a multi-purpose container, consolidating ingredients in the wok before cooking. P3 elucidated: \textit{``I adopt this approach to reduce the complexity and hassle associated with employing multiple containers.''} Similarly, P4 opted to place pepper and other seasonings inside the wok with oil before igniting the flame, effectively condensing multiple steps into a single action, which was typically divided in conventional recipes. Furthermore, our participants exhibited resourcefulness by \textbf{utilizing objects with dual functionalities to diminish the need for additional tools}. For instance, P11 employed a large bowl both as a wok lid and as a receptacle for washing vegetables. P11 explained the practicality of this approach: \textit{``Utilizing the large bowl as a wok lid not only eliminates the necessity for a separate lid but also serves as a convenient vessel for washing various vegetables.''}

\subsubsection{\red{Reducing the Need for Precise Movement}}
In their quest for enhanced convenience and reduced demand for precision and accuracy, our participants expressed a preference for \textbf{minimizing complex 3D spatial actions} \textbf{[Orientation]}. They identified strategies that streamlined their actions, such as adjusting their knife-handling technique and adopting a pragmatic approach to discarding waste. Several participants indicated a desire for simplified knife handling, opting for a 2D movement approach by gripping the knife's back edge rather than the blade's tip (P4). This technique allowed for more straightforward and manageable motion when manipulating the knife during culinary tasks. 

To further simplify their kitchen activities, our participants highlighted the practice of initially disposing of waste items in the sink (P6) \textbf{[Orientation]}. This approach was especially advantageous for items that required precise disposal, as participants found it challenging to accurately target a conventional waste bin placed at ground level. By contrast, the sink offered a more accessible, waist-level receptacle, reducing the need for precision and diminishing concerns about transferring waste to a traditional trash bin after cooking. P6 explained: \textit{``This would reduce my effort of accurately throwing the garbage inside the bin...The water sink is big enough and at my waist level so I can easily throw things in it without much effort and worry about transferring them into the trash bin later after cooking.''}


\section{Findings: Contextual Information Communication and Deployment Considerations (RQ3)}
\label{Findings: Contextual Information Presentation and Deployment Considerations (RQ3)}
In this section, we show our findings about contextual information presentation and deployment considerations by people with vision impairments during cooking activities. \red{We present participants' preferences for information granularity in communication (Section \ref{information granularity}), communication modality (Section \ref{communication modality}), and form factor considerations for future technologies (Section \ref{deployment and form factor considerations})}.


\subsection{\red{Preferred Information Granularity and Level of Detail}}
\label{information granularity}
\subsubsection{\red{Providing precise spatial descriptions}}
\label{position reference}
\red{Participants preferred specific, stationary spatial references to describe object positions, using objects in the kitchen as landmarks}. P4 elaborated:

\begin{quote}
    ``I typically say that the salt is on the left-hand side of the gas range, or the oyster sauce is on the windowsill. I prefer not to know if my oyster sauce is close to my sugar, because other people might place it in different spots.''
\end{quote}

\red{Moreover, participants requested detailed spatial references. P1, for instance, suggested describing an object as being on the third shelf of the fridge door, rather than simply mentioning that it is on a shelf. Best practices here depend on layout: a kitchen with only window has objects ``on the windowsill'', while more details are needed for a kitchen with multiple windows. In some cases, participants combined multiple objects to describe a location,} such as \textit{``the bowl is inside the cabinet at the left-hand side of my gas range.'' (P1)}.

\subsubsection{\red{Limiting verbosity}}
\label{performance and verbosity}
\red{Participants expressed a dislike for systems that chatter constantly, as it could distract them from tasks that require focused auditory attention.} P2 conveyed this sentiment:

\begin{quote}
    ``I don't want the system to constantly provide verbal updates for every new object it encounters. It should provide contextual information thoughtfully.''
\end{quote}


\red{Participants requested that notifications be provided when most needed, and taking into account the user's current location} (P4, P7, P10). P7 elaborated:

\begin{quote}
    ``I don't require precise navigation to a specific reference point in my kitchen, such as the position of my sink. I can manage that on my own. What I need is a brief notification to confirm that I've reached the desired position.''
\end{quote}

\subsection{\red{Preferred Communication Modalities}}
\label{communication modality}
\subsubsection{Combining Speech and Non-Speech Audio Feedback}
\label{Presentation of Contextual Information}

During the post-interviews, we asked participants about their preferred modes of receiving information in the kitchen. A majority of participants preferred non-speech audio notifications, such as a familiar "beep-beep-beep" sound, as these would be less distracting. This preference was primarily motivated by the desire to minimize cognitive load during cooking, as noted in Section \ref{sound reliance}. P4, P8, and P10 mentioned the advantage of less distracting audio feedback, particularly when engaged in conversation or consuming other auditory content. P10 pointed out:

\begin{quote}
    ``I prefer non-verbal sounds over human voice notifications as they are less disruptive while I'm cooking. If I'm in the middle of cooking and listening to a lecture, having a human voice as a notification can disrupt my experience.''
\end{quote}

\red{Participants noted that speech was still needed to convey details, but they requested control over when speech feedback was offered.} 
As articulated by P11:

\begin{quote}
    ``I wouldn't want the system to speak to me in a human voice unless I specifically request it. Typically, not during active cooking, but perhaps I might ask for more detailed information before or after I perform certain actions, such as inquiring about the quality of the dishes I've prepared.''
\end{quote}

Overall, participants hoped for a balance between less distracting audio cues and on-demand speech feedback.

\subsubsection{\red{Offering Proactive Reminders}}
\red{Furthermore, our participants desired proactive reminders around key topics such as safety- and health-related information (P8, P11), as noted} in Sections \ref{safety-related information} and \ref{health-related information}. \red{Participants stressed the importance of being able to edit and customize reminders. Reminders could be triggered by events in the kitchen or based on the user's location.} P12 elaborated on this concept, likening it to an indoor navigation system:

\begin{quote}
    ``This would resemble an indoor navigation system that can automatically provide me with the references I've designated as relevant to my current location.''
\end{quote}

\subsubsection{\red{Increasing Environmental Awareness}}
\label{tracking opportunities}
\red{When discussing the benefits of using a wearable camera to track the user's activities in the kitchen, nine participants emphasized that such a system would be tolerable, but that they also might need awareness of events happening outside the view of a wearable camera}. P5 articulated this need:

\begin{quote}
    ``I'm unsure if I can always have the tracking system facing the exact objects or reference points. Therefore, I want the system to be capable of providing me with contextual information even when I'm not directly facing it.''
\end{quote}

\subsection{\red{Deploying New Technologies in the Kitchen}}
\label{deployment and form factor considerations}
\subsubsection{\red{Kitchen Deployment Considerations}}
\red{During the contextual inquiry, we inquired about our participants' interest and concerns related to technology that could track contextual information.} Our participants raised \red{concerns related to introducing new technology into cooking}, including the risk of water splatter, exposure to fire, oil-proofing, battery life, and dirt resistance. \red{P9 also expressed concerns about the potential high cost of such a system, suggesting that additional support could be provided in existing mobile device hardware.}

\subsubsection{\red{Device Form Factors and Location}}
\red{Assuming new technology became available to support activities in the kitchen, we asked participants whether} they would prefer a stationary system installed in the kitchen or a body-worn system. Out of the 12 participants, 11 expressed a preference for body-worn systems due to factors such as ease of installation, extensibility, and cost efficiency. P4 raised questions about the challenges of installing technology themselves. P7 mentioned that new technology might also be used for other purposes, such as indoor navigation. One participant favored a stationary camera due to privacy concerns associated with wearing such a system during other activities (P12).

\red{In addition to this feedback}, we conducted an evaluation in which participants wore the camera in various body positions, including the chest, head, and wrist, \red{and asked about their preferences for device location.}. Eight participants indicated a preference for a chest-worn camera, while four favored a head-worn camera. Those in favor of the chest-worn position cited its non-obtrusive nature, expressing concern that a head- or hand-worn device might  collide with obstacles (P5). Participants who preferred the head-worn position emphasized the advantages of greater freedom of movement when engaged in cooking tasks.

\section{Discussion and Future Directions}
\label{Discussion}
\red{In this section, we explore how the contextual needs identified in this study might be integrated into future technology such as AI-powered kitchen assistants.}

\subsection{\red{Importance and Opportunities for AI to Improve Contextual Awareness}}
Prior research has explored various ways of adopting AI-powered systems to identify objects, such as using computer vision with overhead RGB-D cameras \cite{stein2013combining} or recognizing objects in mobile device images \cite{SeeingAI86:online,Introduc36:online}. \red{While these systems typically detect objects, our study notes the importance of both identifying objects and describing the contextual information of an object. For example, existing systems mostly identify the object name (e.g., milk bottle) or a group name of objects (e.g., vegetables), instead of providing more contextual information to people (e.g., distance to the object, expiration date of food). Based on our findings on what and how contextual information should be presented, there is also an opportunity for these AI-powered systems to provide more precise spatial descriptions that are customized to the user's workspace. Beyond kitchen contexts specifically, we recommend future research to also explore other scenarios that embedding specific contextual information of objects can support people with vision impairments with agency and autonomy (e.g., art museum \cite{li2023understanding}, grocery store \cite{zhao2016cuesee}, makeup \cite{li2022feels}).}


\subsection{\red{Creating Smart Objects in the Kitchen}}
\red{In our study, we found that visually impaired participants often substituted touch for visual information, and did additional work to make recognition by touch easier, such as using different bottle shapes for different ingredients (Section \ref{customized creation of contextual information}). If users are already choosing or augmenting the shapes of objects, it might be feasible to expect users to attach tags to objects that could improve recognition \cite{li2021non,guo2017facade}. Using more complex tagging methods such as 3D printed models \cite{shi2017designing} or sensor-enabled tags \cite{suzuki2017fluxmarker} could provide additional contextual information. This leads to future fabrication research to consider exploring: 1) tagging methods that are sustainable and deformable so that they can be attached to different kitchen objects, 2) platforms that support people with vision impairments to create customized 3D objects for their home, 3) low-cost embedded systems to track and report back about item status.}


\subsection{\red{Augmenting the Kitchen or the User}}
\red{Beyond customizing contextual information of objects, we also uncovered the importance of managing the space (Section \ref{spatial information}) and knowing the internal state of objects (Section \ref{status of objects}), which often require them to leverage touch to scan through the space (Section \ref{touch}). Prior research has explored approaches to \textbf{augment the kitchen space}, such as by instrumenting a kitchen with cameras, microphones, and motion tracking \cite{olivier2009ambient,vzaric2021design}. These approaches should also work for people with vision impairments, and may provide even more benefit as they may help address accessibility challenges in the kitchen. Creating and deploying such systems would need to acknowledge that visually impaired users may move and act differently within the kitchen space, and may have particular concerns around issues such as spilling ingredients or tracking cooking status.}

\red{Along with augmenting the kitchen space, most of our participants were open to the idea of using \textbf{wearable devices} (e.g., body-worn cameras or smartwatches) on themselves to track their activities and provide contextually-relevant suggestions (Section \ref{deployment and form factor considerations}). For example, we found that tracking the internal state of the space and objects can be highly visual and it brings opportunities to create systems to visually check the internal state of objects through user definitions (e.g., dumplings floating indicate doneness). To understand visual information, devices with worn cameras could leverage pre-trained Vision-Language Models (VLMs) \cite{gao2023physically,driess2023palm,zhou2022conditional,chao2018learning} to track the status of objects and answer user questions (VQA), such as safety-related questions as well as relative positions and state of objects (e.g., material, quantity) \cite{gao2023physically}. Such a system could also provide proactive notifications, such as noting if an object has moved or if the space has been rearranged by another user of that kitchen.} 


\subsection{\red{Tracking Activities in the Background}}
\red{Participants noted that they would be interested in knowing the status of the kitchen even when they had moved outside of that space (Section \ref{tracking opportunities}). Implicit tracking systems \cite{hodges2011sensecam,karim2006exploiting,gouveia2013footprint} could discreetly and continuously monitor contextual information, such as the location of objects, while the user is engaged in various activities within the kitchen \cite{li2019fmt}. This approach would alleviate the need for users to consciously track objects in the kitchen at all times. Future research should consider 1) achieving comprehensive coverage within diverse kitchen spaces, 2) the practicality of addressing challenges related to power consumption and storage capacity \cite{li2019fmt}, 3) the interactive interface for people with vision impairments to pre-assign object of interests.}


\subsection{\red{Multimodal Interaction in the Kitchen}}
\red{Participants' activities in the kitchen often leveraged multiple sensory modes at once and in concert (e.g., touch, sound, and smell) (Section \ref{multisensory}). Participants also experienced overload and related challenges during these tasks (Section \ref{multisensory}). This type of multimodal interaction is known in HCI to support users with diverse abilities \cite{PreeceRogersSharp15,oviatt2017handbook,liu2023tool,bouchet2004icare,reeves2004guidelines,liu2022wigglite}. As users with vision impairments already interact multimodally in the kitchen, technology that supports these users should also be multimodal and adapted to their existing ways of performing tasks (Section \ref{multisensory}). We noted that participants' abilities to engage the environment were sometimes affected by context, for example, using touch is not always feasible when the user's hands are dirty or greasy, which suggests that future systems should be context-aware and should adapt to the user's activities and current state.}

\section{Limitations}
In our research, our participant pool consisted exclusively of individuals with prior cooking experience. \red{Users with no cooking experience would also benefit from this work, but would likely encounter different problems. All of our participants were either legally or completely blind and did not use any type of vision for assistance while cooking. However, we believe some people with low vision might have different practices and challenges of obtaining objects' contextual information during cooking processes and we suggest future research could further conduct studies with a broader range of visually impaired people on this aspect.} Moreover, it is worth noting that our study participants were recruited locally in mainland China. Consequently, individuals from different geographic regions or diverse cultural backgrounds may exhibit varying behaviors and unique contextual information needs within the kitchen setting \cite{li2021choose}. \red{Extending this work to additional geographic and cultural contexts would likely lead to additional, and complementary, insights.}

\section{Conclusion}
\red{While emerging technologies could support visually impaired users in activities of daily living, it is essential that these technologies are designed with a robust understanding of how users actually perform tasks. Our contextual inquiry study identified the primary sources of contextual information used by visually impaired cooks, and revealed challenges encountered by the visually impaired users when cooking.} Our study \red{also explored preferences and concerns related to the introduction of new technology into the kitchen, and identified preferences for how such technologies should communicate with users (with key issues including mode of communication and levels of verbosity).} Finally, we propose future research directions to increase contextual awareness for individuals with vision impairments \red{when working in the kitchen}. Our work provides initial steps towards enabling individuals with vision impairments to access and track contextual information concerning objects during culinary endeavors.

\begin{acks} 
This work was funded in part by Google Research and NSERC Fellowship PGS-D-557565-2021. We especially thank all participants who accommodated the contextual inquiry in their kitchens. We would also like to sincerely thank Yunfeng Shi and Ziyue Hu for participant recruitment and study support.
\end{acks}

\bibliographystyle{ACM-Reference-Format}
\bibliography{main}


\end{document}